\documentclass[authoryear,preprint,12pt]{elsarticle}

\usepackage{amssymb}
\usepackage{amsthm}
\usepackage{amsmath}

\usepackage{multirow}

\journal{TBA}

\begin{document}

\begin{frontmatter}

\title{Spatial analysis of airborne laser scanning point clouds for predicting forest variables}
\author[1]{Henrike H\"abel}
\author[1]{Andras Balazs}
\author[1]{Mari Myllym\"aki\corref{cor1}}
\cortext[cor1]{Corresponding author: Mari Myllym\"aki, Natural Resources Institute Finland (Luke), Latokartanonkaari 9, FI-00790 Helsinki, Finland; e-mail: mari.myllymaki@luke.fi
}
\address[1]{Natural Resources Institute Finland (Luke), Finland}

\begin{abstract}
With recent developments in remote sensing technologies, plot-level forest resources can be predicted utilizing airborne laser scanning (ALS). 
The prediction is often assisted by mostly vertical summaries of the ALS point clouds. We present a spatial analysis of the point cloud by studying the horizontal distribution of the pulse returns through canopy height models thresholded at different height levels.
The resulting patterns of patches of vegetation and gabs on each layer are summarized to spatial ALS features. We propose new features based on the Euler number, which is the number of patches minus the number of gaps, and the empty-space function, which is a spatial summary function of the gab space. The empty-space function is also used to describe differences in the gab structure between two different layers.
We illustrate usefulness of the proposed spatial features for predicting different forest variables that summarize the spatial structure of forests or their breast height diameter distribution. 
We employ the proposed spatial features, in addition to commonly used features from literature, in the well-known k-nn estimation method to predict the forest variables. We present the methodology on the example of a study site in Central Finland.
\end{abstract}


\end{frontmatter}


\section{Introduction}
\label{sec:introduciton}

Today forest information is collected at many levels and for many purposes, where remote sensing plays an increasing role for forest resource estimation methods \citep{Kangas2018}. Due to imperfect correlation between the remote sensing data and observations collected on field plots, important factors influencing the goodness-of-fit are the plot design, expected number of trees, and the prevailing structure and diversity of the forest \citep{Barrett2016, Tomppo2017}. In this paper, we present how ALS-assisted forest inventories can benefit from spatial pattern analysis and propose spatial summaries of airborne laser scanning (ALS) point clouds that can improve the prediction of forest variables.
\par

In the literature, there are many spatial point cloud summaries describing the patch structure of thresholded canopy height models (CHMs) or measuring the canopy complexity in different ways, see e.g. \citet{Kane2008, Kane2011, Li2014, Zhang2017} and the references therein. The main focus in these studies was to relate canopy complexity to the three-dimensional structure of the forest. Other studies aimed at utilizing remote sensing data to classify the spatial structure of forests determined by the spatial patterns of trees either forming regular, random, or clustered patterns. \citet{Uuttera1998}, for instance, applied individual tree segmentation to high-resolution aerial photographs for this purpose, but experienced difficulties especially with clustered patterns of trees. \citet{Packalen2013} tried a method based on individual tree detection as well, but also, similar to \citet{Pippuri2012}, aimed at classifying and predicting the spatial arrangement of trees with an area based approach for ALS data. This is also the focus of this study on the example of field and ALS data from a study region in Central Finland. Different from \citet{Pippuri2012}, and \citet{Packalen2013}, who predicted these structure classes directly on a categorical scale, we study the spatial structure on a continuous scale.\par

We employ two different groups of ALS point cloud summaries, in short ALS features, for predicting the spatial structure of forest, diameter at breast height ($dbh$) distribution and stand development class, where the latter is also in style of the work by \citet{Pippuri2012}. The first group includes common ALS features following the practice of the management inventory \citep{Tomppo2017}. Since the majority of them summarizes the pulse returns vertically, we shall call them vertical features. The second group is formed by spatial ALS features extracted from thresholded CHMs of which some are introduced for the first time in this context to the best of our knowledge.\par

In order to create a thresholded CHM, \citet{Pippuri2012} set the threshold level to 5 m above ground, where pulse return values below the threshold were declared as gap and those above as canopy patch. \citet{Packalen2013} chose an adaptive threshold based on the maximum height of the CHM, which on average set the threshold to about 70\% of the maximum height. Instead of thresholding the CHM only once, we suggest to use several thresholds at different height levels. We propound to use the Euler number, which is the number of vegetation patches minus the number of gabs, as a simple measure of canopy complexity. Furthermore, the so-called empty-space function from spatial statistics plays a key role in this study since it is used for calculating both spatial ALS features and a forest variable measuring the spatial structure on plot-level.\par

 We concentrate the comparison of our new spatial ALS features to the work by \citet{Packalen2013}, since their study set-up and aims are the closest to ours. We show that including our proposed spatial ALS features adds valuable information in particular to the prediction of the spatial structure of forests.

\section{Materials}
\label{sec:material}

\subsection{Field data}
\label{sec:fielddata}
A total of 2469 field plots was measured on a study site in Central Finland in 2013 following a systematic cluster sampling. The land area was 5700 km$^2$, of which 4310 km$^2$ were forestry land including also poorly productive forest land and unproductive land as defined in the Finnish national forest inventories and management inventories. The topography is relatively flat with elevation values generally between 100 m and 200 m above sea level. Belonging to the southern and middle boreal vegetation zone, the forests are mainly coniferous, where Scots pine (\emph{Pinus sylvestris} L.) and Norway spruce (\emph{Picea abies} [L.] H. Karst.) are the most common species. The principal silvicultural system in the region has been even-aged management \citep{Tomppo2017,Tuominen2017}. \par

All trees with $dbh$ greater than 4.5 cm were measured for fixed radius plots of 9 m (ca.\ 254 m$^2$). More details on all measurements made can be found in \citet{Tomppo2017}.\par

In this study, a subset of plots within single stands with at least 10 trees and available ALS data were considered and only the location, and $dbh$ of tally trees were included in the analysis. This resulted in a total number of 1161 plots and 34965 measured trees. Table \ref{tab:data} summarizes forest characteristics of the selected plots organized according to development class. Most plots (92\%) were classified as either young thinning stands (Class 4), advanced thinning stands (Class 5), or mature stands (Class 6).

\begin{table}[ht]
\centering
\begin{tabular}{lrrrr}
  \hline
& \multicolumn{4}{c}{Development class}\\
                    & 4: young             & 5: advanced             & 6: mature             & other\\  \hline
No. plots           & 305           & 582           & 183           & 91 \\
Mean diameter, cm   & 10.46 (2.08)  & 15.06 (3.18)  & 17.09 (4.42)  & 7.68 (1.66)\\ 
Basal area, m$^2$/ha& 14.24 (6.06)  & 21.09 (6.80)  & 25.90 (9.23)  & 3.68  (1.90)\\
Pine, \%            & 0.67 (0.36)   & 0.56 (0.33)   & 0.37 (0.31)   & 0.72 (0.40)\\
Spruce, \%          & 0.14 (0.23)   & 0.24 (0.26)   & 0.38 (0.31)   & 0.17 (0.32)\\
Broadleaved, \%     & 0.19 (0.26)   & 0.20 (0.20)   & 0.24 (0.24)   & 0.11 (0.25)\\
No. of stems/ha     & 1544 (701)    & 1128 (575)    & 999 (515)     & 711 (249) \\ 
   \hline
\end{tabular}
\caption{Average values (standard deviations) of the forest variables calculated from trees with $dbh \geq 4.5$ cm on 1161 plots summarized for development classes young thinning stands (4), advanced thinning stands (5), mature stands (6), and others (regeneration and seedling stands as well as unknown). The species percentages refer to proportions of basal area per species of the total basal area per plot.}
\label{tab:data}
\end{table}

\subsection{Validation data}
\label{sec:validata}
In the same region in Central Finland, 30 additional plots were measured in 2014 for the purpose of model validation. The plots were of size 32 m $\times$ 32 m and subdivided into four subplots of size 16 m $\times$ 16 m (256 m$^2$). The plots were selected at locations where ALS-based estimation was expected to result in large root mean squared errors \citep{Tomppo2017}. Almost all validation plots were thinning stands, of which 17 were young (Class 4) and 10 were advanced (Class 5). Even though also smaller trees were measured, only trees with a $dbh$ greater than 4.5 were included in a validation study to match the 2013 data.

\subsection{ALS data}
\label{sec:ALSdata}
The ALS data were acquired by Blom Kartta Oy, Finland, for the operative management inventory by the Finnish Forest Centre between 28 June and 27 August 2013. The Piper Navajo airplane and the Optech Gemini ALTM scanner were used with the following parameters: flight altitude 1730 m, strip overlap 20\%, pulse frequency 70,000 Hz, scanning frequency 37 Hz, half scan angle 20 degrees, pulse density 0.89/m$^2$, and maximum number of observed pulse returns 4. For the analysis presented here, only the ALS data at the field plots were used. The extracted ALS feature variables are described in Section \ref{sec:features}.

\section{Methods}
\label{sec:methods}

Prior to introducing the ALS features in Section \ref{sec:features} and the studied forest variables in Section \ref{sec:indices}, Section \ref{sec:summaries} first gives necessary background for the spatial analysis. Section \ref{sec:estimation} explains how the forest variables are predicted for field plots using the ALS data.\par

All computations were conducted with the statistical software R (version 3.4.4.)\ and mainly using the packages \texttt{spatstat} \citep{spatstat}, \texttt{spatialgraphs} \citep{spatgraphs}, and \texttt{lidR} \citep{lidR}.

\subsection{Preliminaries on spatial statistics}
\label{sec:summaries}

In this application, tree locations are mathematically expressed as a point pattern with a finite number of $n$ trees observed on a field plot $W\subset \mathbb{R}^2$. Each observed point pattern is interpreted as a realization of a planar point process, which is assumed to be translation and rotation invariant with intensity $\lambda$. Here, $\lambda$ can be interpreted as the tree density per square meter.\par

A point pattern is called completely spatially random (CSR) if there is no interaction between the points. Comparing to the CSR case, interaction between the points may result in either larger inter-point distances and regular patterns or smaller inter-point distances and clustered patterns. Regularity and clustering may also occur in the same pattern, but at different distances. Due to the small field plot size in this study, distances only up to 4.5 meters were taken into account and, thus, the spatial structure of forests, clustering or regularity, was considered only within this range.\par

\begin{figure}[ht]
	\centering
		\includegraphics[width=.33\textwidth]{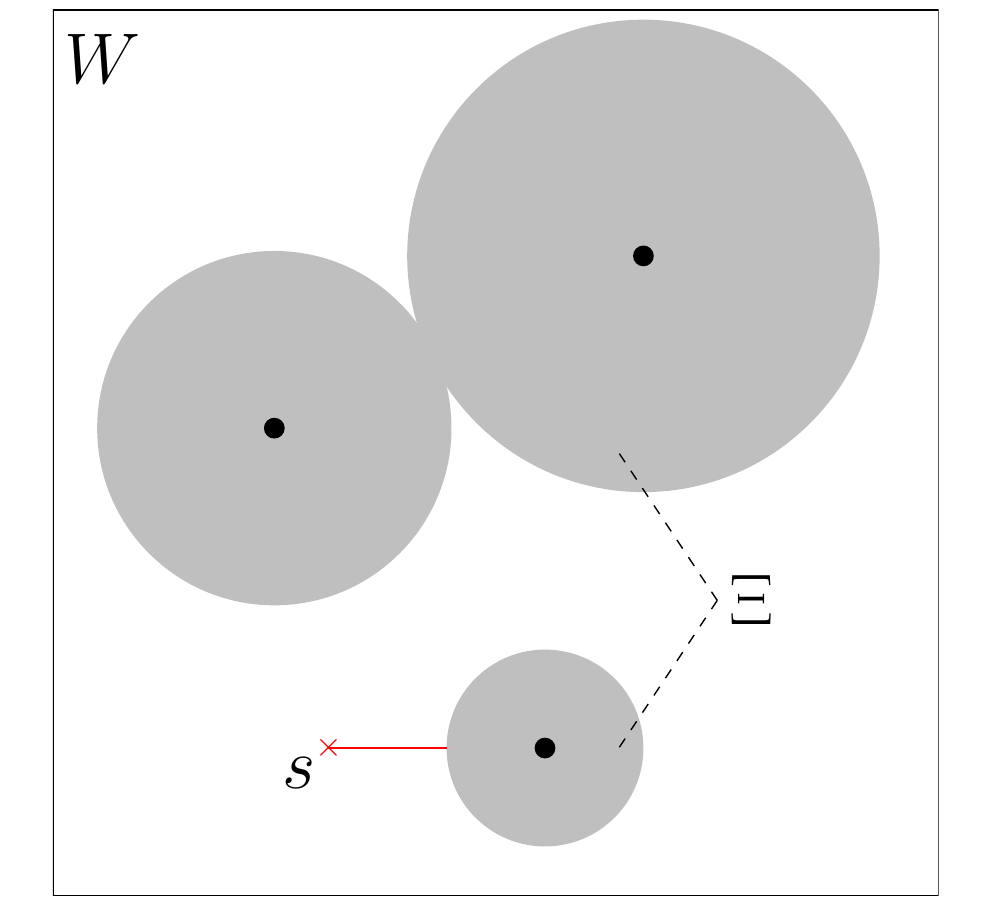}
		\caption{Schematic illustration of the random set $\Xi$ (e.g.\ canopy patches) in an observation window $W$ and an arbitrary location $s$ in $W\setminus \Xi$ and its shortest distance to $\Xi$ (red solid line).}
\label{fig:randomset}		
\end{figure}

 Let us now consider a random set $\Xi$ of discs with a random radius $\mathcal{R}$ centered at random locations forming a point pattern in an observation window $W$. For instance in this application, $\Xi$ consists of the canopy patches. The empty-space function $F$ then gives the cumulative distribution function of the distance from an arbitrary location $s$ in the `empty' space $W\setminus \Xi$ to the nearest point in the random set $\Xi$ (Figure \ref{fig:randomset}). In the case of the Boolean model, which serves as a reference model with discs located uniformly on $W$, a theoretical $F$-function for all distances $r>0$ is given by
\begin{equation}
	\label{eq:F}
	F_{\text{theo}}(r)=1-\exp(-\lambda\pi r(2\mathbb{E}[\mathcal{R}]+r)),
\end{equation} 
where the area fraction of the discs $p=1-\exp(-\lambda\pi\mathbb{E}[\mathcal{R}^2])$
can be used to calculate the expected number of disc centers per unit area ($\lambda$). Thus, $\lambda$ in \eqref{eq:F} can be replaced by $-\log(1-p)/(\pi\mathbb{E}[\mathcal{R}^2])$.\par

The empty-space function $F$ can be defined for a point pattern accordingly. The theoretical $F$-function in the CSR case is $F_{\text{theo}}(r) = 1 - \exp(-\lambda\pi r^2)$ for distances $r\geq 0$. The point density $\lambda$ is usually estimated by the number of points observed in $W$ divided by the area of $W$, i.e.\ $n/|W|$. In order to obtain an unbiased estimate for the empty-space function, the spatial Kaplan-Meier estimator was used to correct for unobserved points outside the observation window $W$ \citep{Baddeley1997}. 

The empty-space function $F$ is also often called the spherical contact distribution function and denoted by $H_s$ \citep[pp.\ 42, 87, 115]{ChiuSKM2013}. 

\subsection{ALS feature variables}
\label{sec:features}

The ALS features included in our study were divided into two groups. The first group includes the vertical features (features 1-62 in Table \ref{tab:features}). The second group is formed by the spatial features extracted from thresholded CHMs (features 63-98 in Table \ref{tab:features}) including new features (features 79-98) based on the Euler number and empty-space function $F$.\par

The CHMs were calculated for 12 m circles around each field plot center using the R package \texttt{lidR} with the default values of the implemented pitfree algorithm in combination with the subcircling tweak \citep{lidR}. Beforehand, first pulse returns with heights smaller than 1.3 m above the ground were set to ground level. The ALS features were determined from the inner 9 m circles covering the respective field plots.

\subsubsection{Definition of spatial ALS features}
\label{sec:spatialALS}

The spatial ALS features are defined based on a thresholded CHM. In particular, each CHM was divided into two regions according to a threshold of $q\cdot hmax$ for $q= 80, 60, \ldots, 20\%$ given the maximum height $hmax$ of the CHM. Values above the threshold formed the canopy patches at height level $q$. Values below the threshold are referred to as gaps or empty space.\par

Following \citet{Packalen2013} and other work on quantifying canopy patch characteristics, we calculated the number of patches (features 63-66 in Table \ref{tab:features}), the average patch size (features 67-70), standard deviation of the patch size (features 71-74), and the average number of pixels in a 4-neighborhood of the same type as the focal pixel (either patch or gap pixel, features 75-78).\par

\begin{figure}[ht]
	\centering
		\includegraphics[width=\textwidth]{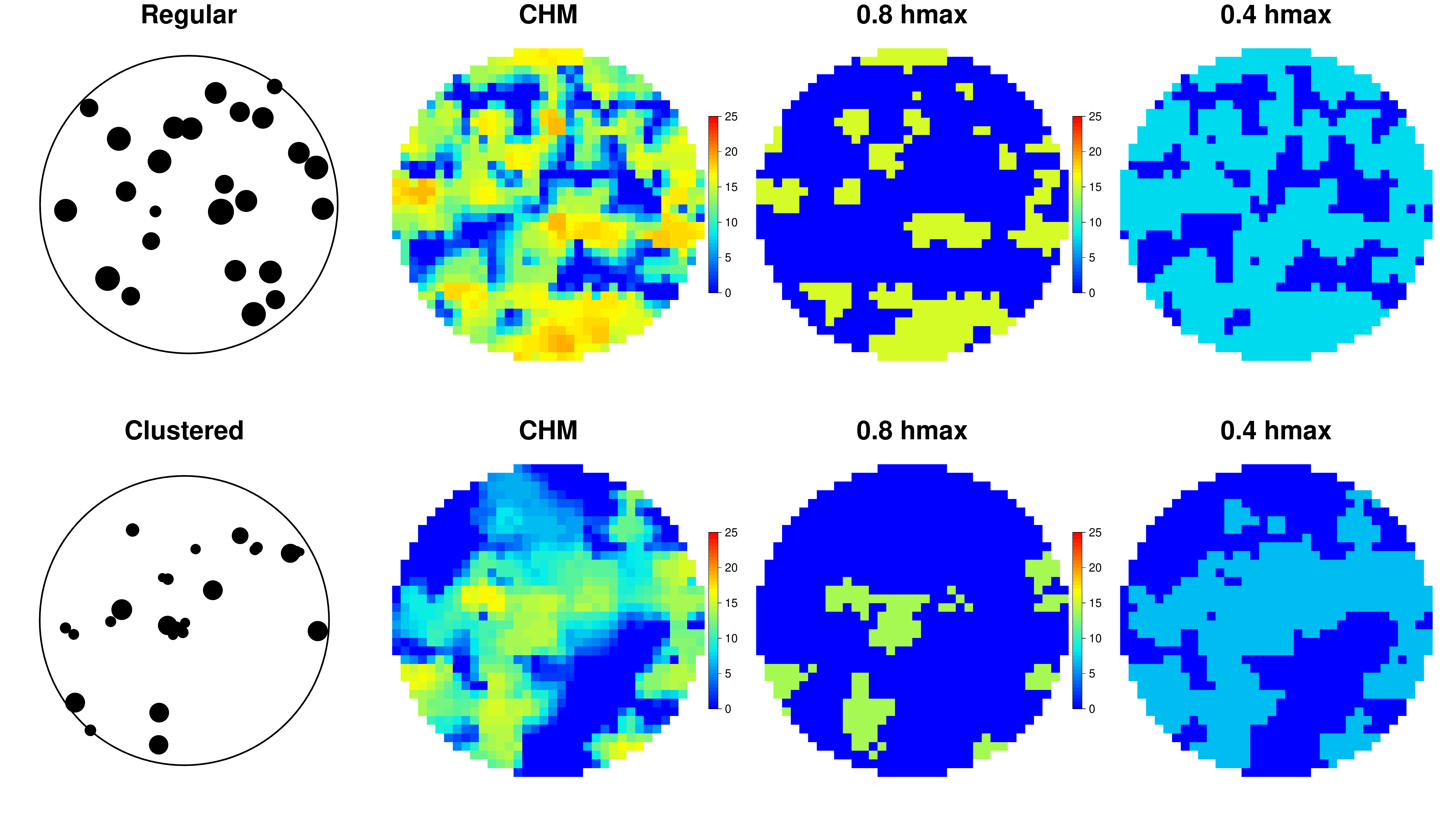}
		\caption{Examples of canopy height models (CHM) and thresholded CHMs for a regular (top row) and a clustered (bottom row) pattern of trees. The thresholds were selected at $80\%$ and $40\%$ of the maximum height (hmax) of each CHM. The points of the patterns of trees on the left have been scaled according to their estimated tree height.}
\label{fig:CHMexample}		
\end{figure}

We additionally suggest the Euler number (features 79-82) to the set of features. It gives the number of canopy patches minus the number of gaps and therewith an easy measure of the canopy complexity. As an illustration for why spatial ALS features at different height levels and including the Euler number are meaningful especially in combination, let us consider the regular and clustered plots presented in Figure \ref{fig:CHMexample}. All plots have approximately the same number of trees (around 25) and almost no gaps at the 80\% height level, but the regular pattern of trees has more canopy patches (15) than the clustered one (8). At the 40\% height level, the largest difference between the two plots can be observed in the Euler number. Due the differences in spatial forest structure, the regular plot has only one canopy patch and shows many gaps resulting in a very low Euler number of -14 whereas there are still 4 canopy patches on the clustered plot and only a few gaps resulting in an Euler number of -3. \par  

In order to also include information about the gaps or empty space, we introduce spatial ALS features based on the empty-space function $F$. For each thresholded CHM, $F$ was estimated as the empirical cumulative distribution function of distances from all empty space pixels to the nearest canopy pixel. The estimator for $F_\text{theo}$ was based on equation \eqref{eq:F}, where the random radius $\mathcal{R}$ was determined by the average of the largest distance between any two pixel of the canopy patches.\par 

The remotely estimated $F$-functions were summarized in two different ways. First, the integrated squared difference
\begin{equation} \label{eq:FI}
	D_I(F, F_\text{theo}) = \operatorname{sgn}\left(\max_{r\leq r_t}|F(r)-F_\text{theo}(r)|\right)\int_0^{r_t}(F(r)-F_\text{theo}(r))^2 dr
\end{equation}
was considered for a chosen upper limit $r_t>0$ (see Section \ref{sec:summaries}). The integral was multiplied by the sign of the maximal difference to $F_\text{theo}$ ($\operatorname{sgn}$) to make a differentiation between regularity and clustering possible. 
Consequently, the larger the absolute value of $D_I$, the larger the difference to the CSR case in terms of space around a random location in the empty space. For a height level $q$, positive and negative signs relate to regular and clustered patterns of trees that have heights larger or equal to $q\cdot hmax$, respectively.\par

Second, we propose an additional summary of the $F$-function based on a Kullback-Leibler-type (KL-type) divergence of the $F$-function and define it as
	\begin{equation} \label{eq:DKL}
		D_{KL}(F\|F_\text{theo})=\int_0^{r_t} F(r)\log\left(\frac{F(r)}{F_\text{theo}(r)}\right)dr,
	\end{equation}
for a chosen upper limit $r_t>0$ of considered distances. $D_{KL}$ is a simpler version of the cumulative Kullback-Leibler information \citep{Crescenzo2015}. Similar to $D_I$, positive values of $D_{KL}$ indicate regularity and negative clustering between the trees with height $\geq q\cdot hmax$, $q = 80, \ldots, 20\%$. However, the obtained values tend to be generally smaller in their absolute value than the values of $D_I$.\par

The summaries $D_I$ and $D_{KL}$ were also used to compare height layers with each other (features 91-98). It can be expected that the higher layers appear more regular than the lower layers, but that the difference is larger for clustered than for regular plots. For instance, for the example in Figure \ref{fig:CHMexample}, $D_{KL}(F^{(q=0.8)}\|F^{(q=0.4)})\approx 40$ for the regular example, but 83 for the clustered plot. This indicates that the upper layer appears more regular than the lower layer in both cases, but that the difference is larger for the clustered plot.

\renewcommand{\arraystretch}{0.8}
\begin{table}[ht] 
\centering
\begin{tabular}{ll}
  \hline
    \multicolumn{2}{l}{\textbf{Vertical features}} \\
		1     & Height of the canopy\\
		2     & Minimum height of first returns \\
		3  	  & Maximum height of first returns \\
		4  	  & Mean height of first returns \\
		5  	  & Standard deviation of heights of first returns \\
		6  	  & Skewness of heights of first returns\\
		7  	  & Kurtosis of heights of first returns \\
		8  	  & Width of range of heights of first returns \\
		9-15  & The features similar to 2-8 for last returns \\
		16 	  & Proportion of canopy returns, all pulse returns \\
		17-27 & Percentiles (5, 10, 20, $\ldots$, 90, 95\%) for first returns\\
		28-38 & Same features as 17-27 for last returns \\
		39-49 & Cumulative proportions of foliage returns 0-5, 5-10, 10-20,
				  $\ldots$, 90-95\%\\
		50-60 & Same features as 39-49 for last returns \\
		61-62 & Mean intensity (first and last returns)\\
       \multicolumn{2}{l}{\textbf{Spatial features at 80, 60, 40, 20\% levels of the maximum height}} \\
        63-66 & Number of patches\\
        67-70 & Average size of patches in number of pixels\\ 
        71-74 & Standard deviation of size of patches\\ 
        75-78 & Average number of same pixel type in a 4-neighbourhood\\
        79-82 & Euler number for TCHMs\\
        83-86 & Integrated deviation of $F$-function from theoretical reference\\
        87-90 & KL-type divergence of $F$-function from theoretical reference\\
        91-94 & Pairwise integrated difference of $F$-functions between TCHMs\\
        95-98 & Pairwise KL-type divergence of $F$-functions between TCHMs\\
   \hline
\end{tabular}
\caption{ALS feature variables calculated on the basis of common summaries of pulse return values (vertical features) and spatial information from thresholded canopy height models (TCHM). The foliage returns were in the range of the height of the first and last returns, respectively.}
\label{tab:features}
\end{table}

\clearpage

\subsection{Forest structure variables}
\label{sec:indices}

The forest variables included in this study are divided into two groups. The first group deals with the spatial structure of forests and the second group with the variation in size.\par

The spatial forest structure has long been quantified by the aggregation index $R$ \citep{Clark1954}. It gives information about the spatial structure of trees with locations $(x_1,\ldots,x_n)$ based on their nearest neighbor $\operatorname{nn}$ and is estimated by
 		\begin{equation} \label{eq:R}
			R = \frac{2}{\sqrt{n|W|}}\sum_{i=1}^n\|x_i-\operatorname{nn}(x_i)\|, 
 		\end{equation}
where $R\approx 1$ in the CSR case, $R>1$ indicates regularity and $R<1$ clustering. In theory, the aggregation index can obtain values between zero and 2.1491. For the plots in Figure \ref{fig:CHMexample}, $R_\text{reg}=1.17$ and $R_\text{clu}=0.76$.\par

In addition to $R$, we use the KL-type divergence \eqref{eq:DKL} of the estimated empty-space function $\widehat{F}$ from its theoretical counterpart $\widehat{F}_\text{theo}$, i.e.
\begin{equation}
    \label{eq:FD}
    FD=D_{KL}(\widehat{F}\|\widehat{F}_\text{theo}).
\end{equation} 
$FD\approx 0$ in the CSR case, $FD>0$ indicates regularity and $FD<0$ clustering. For the plots in Figure \ref{fig:CHMexample}, $FD_\text{reg}=23$ and $FD_\text{clu}=-33$.\par

In order to quantify the variation in size, we assumed that the $dbh$ distribution can be described by a two parameter Weibull-distribution. The scale parameter is related to the $dbh$ range and the shape parameter to the skewness of the distribution \citep{deLima2015}. Small values of the shape parameter can be associated with uneven-aged and large values with even-aged forests. Also the larger the shape parameter, the larger the mean $dbh$ and the smaller the skewness.\par

We further studied the stand development class. The development class is not only related to $dbh$ distribution, but also to basal area, tree density, and management status \citep[Table 2.17]{TomppoEtal2011}. It is the only categorical variable included in this study.

\subsection{Feature selection and prediction of forest variables}
\label{sec:estimation}

Feature selection and prediction of the forest variables were carried out using a genetic algorithm along with the $k$-nearest neighbor method ($k$-nn method) as described by \citet{Tomppo2004} and \citet{Tuominen2017,Tuominen2018} for continuous and discrete variables respectively.\par

In a preceding step, all features were standardized to have the same variation. Then the genetic algorithm implemented in the \texttt{Genalg} package in R \citep{genalg} was used to select the relevant features $f_{l,\cdot},~l=1,\ldots,m$, and to determine the weights $\omega_l$ for them according to a fitness function based on RMSE and absolute bias, which we set to have equal importance. The selection of the optimal $k$ among the tested values (3-6) was included in the routine. The forest variable $y$ of the plot $p$ was predicted using the set of $k$ nearest neighboring plots $I_p$ with $p\notin I_p$: 
\begin{equation} \label{eq:knnReg}
	\widehat{y}_p =\sum_{i\in I_p}w_iy_i
   \quad \text{with} \quad
    w_i = d_i^{-g}\sum_{i\in I_p}d_i^{-g},
\end{equation}
where the weights $w_i$ were determined by the distance between each neighbor $i$ and the plot $p$ in the feature space, namely
\begin{equation} \label{eq:knnDist}
	d^2_{i,p}=\sum_{l=1}^{m}\omega_l^2(f_{l,i}-f_{l,p})^2,
\end{equation}
and the factor $g$.
Different values for $g$ (0-3) were tested, where $g>0$ implies that neighbors with smaller distances get larger weights. The final values for $k$ and $g$ are given in Table \ref{tab:knn}.

\section{Results}
\label{sec:results}

\subsection{Relevance of spatial features}
\label{sec:featureSelect}

The spatial features were sufficiently correlated with the field data, such that four to seven spatial features were selected by the genetic algorithm (Table \ref{tab:knn}). For the spatial structure variables, the spatial features made up 43\% on average of all selected features. For the parameters of the Weibull-distribution for $dbh$, it was 37\% and for plot development class 38\%. The spatial features selected most often were the average number of same pixels in a 4-neighborhood (features 75-78 in Table \ref{tab:features}) and $F$-function based features (features 83-98 in Table \ref{tab:features}). The integrated difference measure $D_I$ in \eqref{eq:FI} was chosen mostly for the comparison to the theoretical $F$-function on the 80, 40, and 20\% height level and the KL-type measure $D_{KL}$ in \eqref{eq:DKL} for the comparison of the $F$-function of two different height levels. Interestingly, the Euler number was only selected for the scale parameter of the Weibull-distribution for $dbh$.\par

Comparing the predicted values with spatial features to predictions made without them, we observed a 6\% reduction of RMSE on average for the spatial structure variables. There was not a prominent improvement for the parameters of the Weibull-distribution for $dbh$, where the RMSE was reduced by 2\% on average. Also the overall accuracy of the development class prediction only improved by 2\%, but Cohen's kappa by 4\%.\par

\begin{table}[ht]
\centering
\begin{tabular}{lcrrrrrr}
  \hline
  Forest      & g & \multicolumn{2}{c}{RMSE} & \multicolumn{2}{c}{Bias} &Spatial &Total \\
  variable      &      & 2013  &	2014 & 2013 & 2014 &  &  \\
  \hline
  $R$        &  0.9 & 0.178  &	0.296  &  0     & -0.136 & 7 & 15\\
  $FD$       &  1.8 &  16.453 &	12.712 &  0     & 2.359  & 7 & 18\\
  Wei. scale &  2.3 &  2.196  &	2.950  &  0.001 & -0.009 & 4 & 15\\
  Wei. shape &  1.1 &   1.1   &	1.299  &  0.001 & 0.448  & 7 & 15\\
  \\         &      & \multicolumn{2}{c}{OA} & \multicolumn{2}{c}{Kappa} \\
  Dev. class & 1.2 & 0.80 & 0.80 & 0.68 & 0.67 & 5 & 13\\
  \hline
\end{tabular}
\caption{Outcome of the genetic algorithm for feature selection and prediction of the spatial structure variables $R$ and $FD$, the scale and shape parameter of the Weibull-distribution for $dbh$, and the development class (Dev. class) for the 2013 data including spatial features and the validation data from 2014. In all cases $k=6$ neighbors were considered, but different scaling  values $g$ were given to the weights. For the continuous variables RMSE and bias are given and for the categorical the overall accuracy (OA) and Cohen's kappa (Kappa). The number of spatial ALS features (Spatial) and the total number of selected features (Total) are listed.}
\label{tab:knn}
\end{table}

\subsection{Prediction of forest variables}
\label{sec:prediction}

\begin{figure}[ht]
	\centering
        \includegraphics[width=\textwidth]{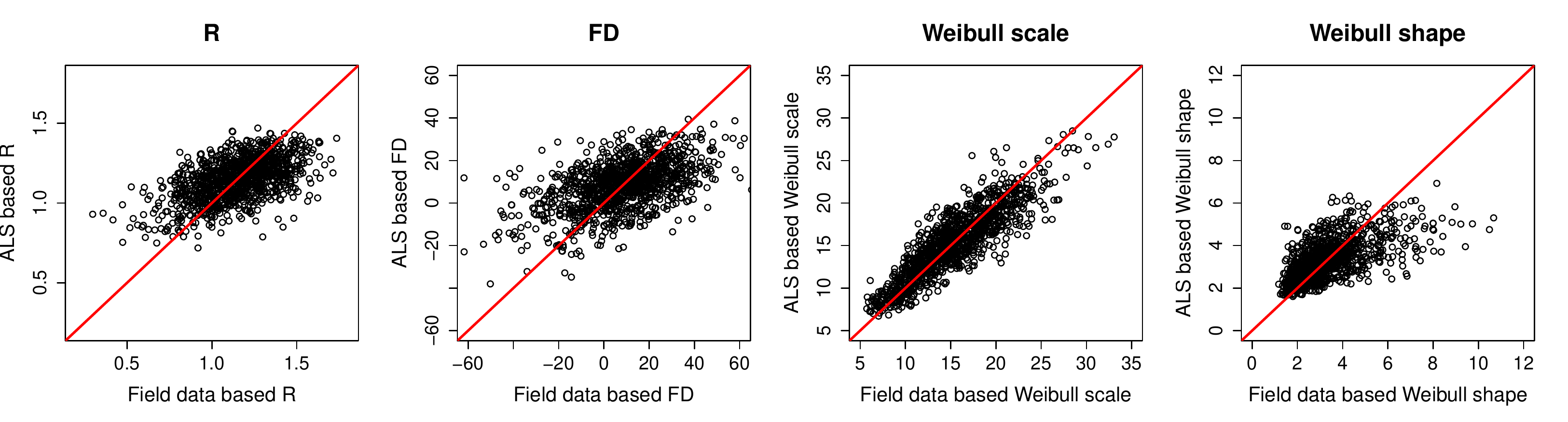}
		\caption{ALS-based forest variable estimates versus field data-based values for the spatial structure variables $R$ and $FD$ and the scale and shape parameter of the Weibull-distribution for $dbh$.}
\label{fig:fit}		
\end{figure}

All estimates of the studied forest variables were practically unbiased (2013 in Table \ref{tab:knn}). The predicted values for $R$ and $FD$ both showed a tendency to be overestimated for negative values and underestimated for positive (Figure \ref{fig:fit}). Consequently, more patterns of trees appear random and fewer forest were classified as regular or clustered based on the ALS data. This trend was more pronounced for $FD$ than $R$. The scale parameter of the Weibull-distribution for $dbh$ associated with the tree size range was predicted well, but the shape parameter related to the skewness of the $dbh$ distribution was underestimated for values larger than six. This means that even-aged forests were predicted to have a skewer $dbh$ distribution (Figure \ref{fig:fit}). Also the development classes were predicted well with an overall accuracy of 80\% and a Cohen's kappa of 0.68 (Table \ref{tab:knn}). The largest error was made for mature stands (Class 6), where 40\% of the plots were missclassified as advanced thinning stands (Class 5). The complete error matrix can be found in Table \ref{tab:classifyDevclass} in the appendix.

\subsection{Classification of spatial structure}
\label{sec:structure}

In a further analysis for the comparison to \citet{Pippuri2012} and \citet{Packalen2013}, the spatial structure variables $R$ and $FD$ were used in classifications of field plots into regular, random, and clustered patterns.
In contrast to these previous works, we used simple cut-off values in a more practical approach. These cut-off values were 0.85 and 1.15 for $R$ and $\pm$ 15 for $FD$.\par

The two field data-based classifications agreed on 342 regular, 310 random, and 50 clustered patterns for the field data, which means a 60\% agreement. A 59\% agreement was achieved with the ALS-based classification, where more plots (434) were jointly classified as random and only 245 as regular and four as clustered patterns. 191 of these ALS-based regular plots were also classified as such with the field data and only two of the latter four plots were also clustered field data-based plots.\par

Table \ref{tab:classify} presents the results of the classification for each spatial structure variable separately. The observed values refer to the values estimated from the field data. The overall accuracy, Cohen's kappa, and differentiation of clustered and regular patterns were the best for $R$. There were difficulties with both variables to clearly distinguish the random patterns. Furthermore, some regular plots were missclassified as clustered and vice versa. This was in particular a problem for the forest variable $R$. One of the reasons why regular plots have been missclassified as clustered seemed to be many relative short distances between trees. Given the low resolution of the ALS data, this led to a patch structure resembling a clustered pattern. The main reason for clustered plots being missclassified as regular plots appeared to be an extremely large variation in tree size, where small trees where covered by large trees resulting in a patch structure of a regular pattern.

\begin{table}[ht!]
\centering
\begin{tabular}{lllrrrr}
  \hline
 &&&\multicolumn{4}{c}{Predicted} \\
 &&&                   Regular & Random & Clustered & Total \\
  \hline
                          &\multirow{4}{*}{$R$} & Regular   & 440 & 176 &   2 & 618  \\ 
                          &                    & Random    & 172 & 264 &   4  & 440  \\ 
                          &                    & Clustered &   10 &  75 &  18  & 103 \\ 
                          &                    & Total     &622 &515 &  24  & 1161\\ 
Observed & & & & & \\ 
                          &\multirow{4}{*}{$FD$} & Regular   & 183 & 246 &   0  &  429   \\ 
                          &                    & Random    &  111 & 480 &  15  & 606   \\ 
                          &                    & Clustered &   4 &  97 &  25  &  126 \\ 
                          &                    & Total     & 298 & 823  & 40  & 1161\\
   \hline
\end{tabular}
\caption{Error matrices for prediction of spatial structure by classifications based on the aggregation index $R$ and the empty-space function $F$ based summary $FD$. The overall accuracy was 62.2\% for $R$ and 59.3\% for $FD$ with a Cohen's kappa of 0.31 and 0.23, respectively.}
\label{tab:classify}
\end{table}

\subsection{Validation study}
\label{sec:validation}

The data from 2014 (see Section \ref{sec:validata}) was used in a validation study. The features and weights selected for the 1161 circular plots measured in 2013 were used to predict the forest variables of the 120 subplots measured in 2014 by the $k$-nn method, where only the 2013 plots were allowed as possible neighbors for the 2014 subplots. In this way, any issues due to correlations between subplots from the same 32 m $\times$ 32 m plot were avoided.\par

A goodness-of-fit analysis showed that the RMSEs were in about the same range for validation study compared to the main study with the 2013 data, but the absolute bias increased (2014 in Table \ref{tab:knn}). The prediction of the development class only had a sightly lower Cohen's kappa of 0.67.\par

The predicted spatial structure variables were analyzed further with regard to classifying the spatial structure of trees. With the same rules as applied to the 2013 data, the overall accuracy was 55.8\% for $R$ and 68.3\% for $FD$. The values for Cohen's kappa were 0.07 and 0.01, respectively. The ALS-based $R$ prediction classified 52 out of 84 (62\%) regular plots correctly, but failed to detect any clustered patterns. With the ALS-based $FD$ only four regular and no clustered plots were classified correctly, but 78 out of 94 (83\%) random plots were in agreement with the field data classification.\par

Classifying the original 32 m $\times$ 32 m field plots with $R$ resulted in 16 regular, 13 random, and one clustered plot, whereas using $FD$ gave 19 regular and 11 clustered patterns. In an additional significance test for complete spatial randomness with the global envelope test based on the $F$- and Ripley's $K$-function \citep{Myllymaki2017, Mrkvicka2017}, the CSR hypothesis was rejected for all plots leading to the same classification as with $FD$. Testing the 16 m $\times$ 16 m subplots, however, 69 regular, 42 random patterns, and 9 clustered patterns were obtained. These classification and test results lead to two conclusions. First, the quality of the differentiation of regular and especially clustered patterns from random patterns depends on the field plot size. Second, using the aggregation index $R$ is not a good forest variable for detecting clustered forests.

\section{Discussion}
\label{sec:discussion}

This study has confirmed that among the spatial ALS features the ones describing the canopy complexity are the most informative. Introducing the empty-space function to the set of features proofed to be a relevant addition to the 4-neighborhood based summary inspired by \citet{Packalen2013} (features 75-78 in our Table \ref{tab:features}) as both were selected by the genetic algorithm used in the forest variable prediction. In a side study not presented here, we implemented the original set of features by \citet{Packalen2013} (their Table 2), but found that merging the neighborhood counts for patch and gap pixels into one feature led to smaller RMSEs. We also found that the Euler number was selected more often, together with the number of patches (features 67-70), if features 75-78 were not included. Consequently, the Euler number in combination with the number of patches appeared to contain valuable information for the prediction of the studied forest variables, but the same information was apparently better captured by the the 4-neighborhood based features.\par

The spatial ALS feature variables were selected by the genetic algorithm for the prediction of all studied forest variables. However, while they improved the prediction of the spatial structure variables, the improvements for other variables were rather small. The aim of this study was to predict the degrees of clustering and regularity, whereby we considered the continuous variables $R$ and $FD$, but originally also further variables based on other spatial summary functions. However, it appeared that this task was a bit too ambitious given the relatively small field plots. Larger plots should result in smaller RMSEs and bias. This leads back to the question of what is the optimal plot design in ALS-assisted forest inventories. It appears that their size should preferably be at least 256 m$^2$ if aiming at prediction of forest structure variables similar to those considered here. In order to keep the costs at a comparable level, designs also other than fixed radius plots should be considered \citep[see e.g.][]{Tomppo2017}.\par

\citet{Packalen2013} used the so-called $L$-function in the classification rule of the field data, but in a computationally heavy Monte Carlo method. The resolution of their ALS data was slightly lower, but still comparable, and they used a smaller number (79 in total) of larger plots of size 20 m $\times$ 20 m and 30 m $\times$ 30 m.  We used simple and practical classifiers based on distance measurements rather than tree locations. Still, we obtained comparable results with those obtained by the AREA method in \citet{Packalen2013} for the 2013 data with the aggregation index $R$. \par
 
 \citet{Pippuri2012} also had slightly lower resolution ALS data and chose the so-called (spatial) $t_N$-index for the classification of 28 microstands in Southern Finland of sizes between 0.2 and 0.7 ha. They achieved better classification results, however, this could be due to the large size of their field plots and greater differences in tree density between regular and clustered patterns compared to our study.\par
 
We had not expected that the aggregation index $R$ would perform well in the classification of the 2013 data as it is a rather simple summary of the spatial distribution. In fact, this simplicity might facilitate its prediction, especially for small field plots. The validation study has shown, however, that even though $R$ was predicted more accurately, it may not be the best classifier for the spatial structure of trees in all possible forest scenarios especially with clustered patterns of trees. Therefore, we expect that better results may be obtained with $FD$ or other summaries for larger field plots. Our study showed that quantifying clustering and regularity of forests from small field plots is itself a difficult task. It is not only statistically difficult to separate clustered and regular forests from CSR based on a pattern of only a few trees, but also different indices can lead to different classifications.

\section{Conclusion}
\label{sec:conclusion}

New spatial ALS features were found for the prediction of forest structure variables in ALS-assisted inventories. Their application was focused on the prediction and classification of the spatial structure of forests, but also variables capturing the $dbh$ distribution and stand development class were considered. We can recommend to include spatial ALS features from CHMs thresholded at more than one height level. Most informative were relatively complex features that describe the spatial structure of the gaps with the empty-space function or that take the pixel type in a 4-neighborhood into account. The empty-space function appears to be especially useful as a measure of differences in the gab structure at two different height levels. A simple alternative to describe the canopy complexity is offered by the Euler number, which is the number of vegetation patches minus the number of gaps.\par

This study has shown the potential of spatial analysis of ALS point clouds through CHMs thresholded at several height levels. Further studies with higher resolution data or larger sample plots could also investigate spatial ALS features based on other spatial summary functions, for instance, morphological functions. Furthermore, it might be worth studying how the spatial structure classification can be improved even further given that the predictions should be more accurate. \par

In conclusion, the presented methodology for spatial ALS features is simple, practical and general enough to be applied to other forest variables such as conventional forest inventory attributes and different three-dimensional remote sensing scenarios.

\section*{Acknowledgements}

Henrike H\"abel and Andras Balazs have been financially supported by the Academy of Finland (Project Number 304212) and Mari Myllym\"aki similarly by the Academy of Finland (Project Numbers 295100 and 306875). The Finnish Forest Centre provided the ALS data. The authors would like to thank Kai M\"akisara for assisting with preparing the data for the study, and Jussi Peuhkurinen (Oy Arbonaut Ltd) and Annika Kangas together with other colleagues at Luke for valuable discussions.

  \bibliographystyle{elsarticle-harv}
 \bibliography{main.bbl}

\clearpage
\appendix
\section{Error matrix for prediction of development class}
\label{app:A}
\begin{table}[ht]
\centering
\begin{tabular}{ccrrrrrrrrr}
  \hline
&  & \multicolumn{9}{c}{Predicted} \\
& & NA & 1 & 2 & 3 & 4 & 5 & 6 & 7 & Total\\ 
  \hline
 \multirow{9}{*}{Observed} & NA & 24 &0&0&   3 &   9 &   0 &   0 & 0 & 36\\ 
  &1 &   0 &   0 &   0 &   0 &   0 &   0 &   0 & 0 & 0\\ 
  &2 &   0 &   0 &   0 &   0 &   0 &   0 &   0 & 0 & 0\\ 
  &3 &   5 &   0 &   0 &  38 &  10 &   0 &   0 & 0 & 53\\ 
  &4 &   4 &   0 &   0 &   7 & 231 &  61 &   2 & 0 & 305\\ 
  &5 &   0 &   0 &   0 &   0 &  26 & 528 &  28 & 0 & 582\\ 
  &6 &   0 &   0 &   0 &   0 &   4 &  73 & 106 & 0 & 183\\ 
  &7 &   0 &   0 &   0 &   0 &   0&   1 &   1 & 0 & 2\\
  &Total &  33 &   0&   0&  48&  280&  663&  137& 0 & 1161 \\ 
   \hline
\end{tabular}
\caption{Error matrix for prediction of development class, where NA indicated missing information.}
\label{tab:classifyDevclass}
\end{table}

\end{document}